\documentclass[amssymb,twocolumn,
aps,floatfix,showpacs]{revtex4}
\usepackage{graphicx}

\begin{document}
\title{Time-dependent R-matrix theory applied to two-photon double ionization of He}
\author{H. W. van der Hart}
\affiliation{Centre for Theoretical Atomic, Molecular and Optical Physics, School of Mathematics and Physics,
Queen's University Belfast, Belfast, BT7 1NN, United Kingdom}

\date{\today}

\begin{abstract} 
We introduce a time-dependent R-matrix theory generalised to describe double
ionization processes. The method is used to investigate two-photon double ionization of He
by intense XUV laser radiation. We combine a detailed B-spline-based wavefunction description in a
extended inner region with a single-electron outer region containing channels representing both
single ionization and double ionization. A comparison of wavefunction densities for different box
sizes demonstrates that the flow between the two regions is described with excellent accuracy.
The obtained two-photon double ionization cross sections are in excellent agreement with
other cross sections available. Compared to calculations fully contained within a finite inner region,
the present calculations can be propagated over the time it takes the slowest electron to
reach the boundary. 
\end{abstract}

\pacs{32.80.Rm, 31.15.A-, 32.80.Fb}

\maketitle

\section{Introduction}

In the last decade, free-electron laser facilities operating in the XUV and X-ray
regimes have become available for scientific investigations \cite{Feld13, Bos13, Yab13}. These facilities
have opened up the research area of laser-matter interactions in intense high-frequency
fields. When a high-frequency laser field interacts with an atom, electrons from many different
shells can be removed, and hence the response of many electrons in the atom needs to be considered.
For an intense laser field, absorption of multiple photons can occur, so that a single laser shot
is capable of ejecting many electrons from many different shells, leading to the creation
of highly charged ions \cite{Ric09,Rud12}. 

This development in laser technology needs to be replicated in theory by developing methods
capable of providing detailed understanding of this new area of laser-matter interactions. For example,
great progress has been made in the theoretical description of sequential ionization processes \cite{Kar13},
so that accurate predictions can be made for the observed ionization stages. However, for other
processes theoretical understanding has been developed to a much lesser degree.

One of the processes, for which there is an urgent need for theoretical code development, is the
description
of non-sequential double ionization processes for general multi-electron
atoms in intense laser fields. One of the
prototypical problems for theoretical investigation is two-photon double ionization of
He at photon energies between 40 and 50 eV, which has been investigated through a wide variety
of methods (see, for just a small subset of methods, \cite{Col02,Fen03,Fei08,Gua09,Pal10,Nep10,Arg13}).
Initial studies of this process obtained two-photon double ionization cross sections which varied by
over one order of magnitude. More recently, it has been concluded, however, that accurate
cross sections
can be obtained using either exterior complex scaling (ECS) or through projection onto Coulomb
functions, if the wavefunction is propagated for sufficiently long times and sufficiently long
distances \cite{Arg13}.
Although significant effort has been devoted to describe multiphoton double ionization processes in helium,
the accurate description of double photoionization processes for general multi-electron atoms from
first principles is still
in its infancy, even after absorption of only a single photon \cite{Yip13}.

Over the last seven years, we have explored the application of R-matrix theory for general multi-electron
atoms to time-dependent processes in intense light fields \cite{Har07,Nik08,Lys09,Lys11,Moo11}. These approaches
have demonstrated that time-dependent R-matrix theory has the capability to accurately describe
ultra-fast dynamics, including correlated multi-electron dynamics \cite{Lys09b} and multi-channel dynamics
\cite{Moo11b,Bro12}, accurately. However, many processes involving ultra-short high-frequency light
pulses involve the removal of two electrons, either sequentially or non-sequentially.
In order to be able to apply time-dependent R-matrix theory to these processes, it is
necessary to provide the approach with the capability to describe double ionization processes. In
the present report, we demonstrate this capability by applying time-dependent
R-matrix theory to study non-sequential two-photon double ionization of He.

\section{Methods}

Of the time-dependent R-matrix approaches available, we apply its most recent version, R-matrix
theory including time-dependence (RMT) \cite{Lys11,Moo11} to study two-photon double ionization of
He. The RMT approach provides better performance for large-scale problems than the previous
implementation of time-dependent R-matrix theory, as it can more efficiently exploit massively
parallel computers. However, so far it has only been applied in the determination of time
delays in Ne photoionization at high intensities \cite{Moo11b}.

The RMT approach adopts the standard R-matrix technique of separating space into two
distinct regions, an inner region in which all interactions between all electrons are taken into account,
and an outer region, in  which exchange effects between the electrons are assumed to be negligible
\cite{Bur11}. In
the inner region, the He wavefunction is described using a large CI expansion involving R-matrix basis
functions confined to a sphere of radius $a$. In the outer region the wave function is described in
terms of a combination of a residual-ion state, which remains confined to the sphere of radius $a$,
and a wavefunction associated with the outer electron, which has now moved beyond this sphere. The approach
to describe this outer electron depends on the nature of the problem. For the current time-dependent
approach, the outer electron is described on an extensive radial grid.

The key procedure in the RMT approach is to establish a highly accurate connection between the wavefunction
in the inner region and the wavefunction in the outer region. The RMT approach adopts a procedure 
to connect these wavefunctions, which differs significantly from other R-matrix approaches
\cite{Moo11,Lys11}.
To connect the inner region wavefunction to the outer region, the outer-region grid is expanded into
the inner region, and the inner region wavefunction is evaluated on this expanded grid. The propagation of
the outer region wavefunction, including the kinetic-energy terms, can then proceed entirely on this grid.
To connect the outer region to the inner region, time derivatives of the outer region wavefunction are
determined at the boundary of the inner region. In the inner region, we then determine the time
propagation not only of the initial wavefunctions but also of these time derivatives. The final
inner-region wavefunction is then built from all these terms. A detailed description of the approach is
given in \cite{Moo11,Lys11}.

In the R-matrix inner region, He is described by a wavefunction expansion in terms of products of box-based
He$^+$ eigenfunctions. These eigenfunctions are, in turn, expressed in terms of B-splines. This
expansion is similar to the expansion used for the description of electron-impact excitation of H \cite{Har97},
and it is also similar to the expansions used by Guan \textit{et al} \cite{Gua09} and Nepstad \textit{et al}
\cite{Nep10} in their investigations of two-photon double ionization of He. However, in the present
calculations, all He$^+$ eigenfunctions in the inner region are from the outset defined as
continuum functions. They are eigenfunctions of $\hat{H}+\hat{L}_b$, where $\hat{H}$ is the
field-free Hamiltonian and $\hat{L}_b$ is the Bloch operator, so that $\hat{H}+\hat{L}_b$ is
Hermititan within the inner region. All He$^+$ eigenfunctions will therefore have a
non-vanishing amplitude and a non-vanishing first derivative at the boundary. Basis functions with
non-vanishing amplitudes at the boundary for both electrons are also used in the intermediate-energy
R-matrix approach, which has recently been employed to investigate double photoionization \cite{Sco12}.

The non-vanishing boundary amplitudes for the He$^+$ eigenfunctions affect the subsequent calculations.
In order to link the outer region wavefunction to the inner region wavefunction, we need to know the
boundary amplitudes of the inner-region He eigenfunctions. To obtain the inner region wavefunction near
the boundary on the expanded outer-region grid, we need to determine how each He eigenfunction in the inner
region connects to each He$^+$ eigenstate at the boundary. Since both electrons are treated as potential
continuum electrons, we take both electrons into account in the determination of the boundary
amplitudes and the wavefunction in the inner region.

To reduce the size of the calculations, we limit the He basis: product basis functions, in
which both electrons have both a field-free energy exceeding 102.2 eV and a small boundary amplitude,
are excluded from the calculations. The angular momentum of the first electron is restricted to
$\ell_{max} = 2$, and we only include total angular momentum up to $L_{max}=3$. Most calculations are
carried out for an inner region size of 35 $a_0$, although additional calculations were carried out for
a box size of 50 $a_0$. In the former box, we use 48 B-splines of order 7 with a mixed
exponential-linear distribution of knot points. In the latter box, we use 68 B-splines.

The laser field is described by a 14-cycle pulse, with a 5-cycle $\sin^2$ turn-on, 4 cycles at peak
intensity of 10$^{14}$ W/cm$^2$, and a 5-cycle $\sin^2$ turn-off. For a photon energy of 42.2 eV,
additional calculations were carried by varying the number of cycles at peak intensity between 0 and
4. Within RMT theory, the laser field is described in the length gauge.
The wavefunction is propagated for a time corresponding to 15 cycles of the laser field. Time propagation
is achieved through an Arnoldi propagator of order 8 with a time step size of approximately 0.005 au \cite{Smy98}.
The
grid spacing in the outer region grid is 0.08 au. The total number of channels included in the outer
region is 744 for a box size of 35 $a_0$. The programme runs over 3072 processors using both MPI and OpenMP parallelisation. 

\section{Results}

\begin{figure}
\includegraphics[width=8.5cm]{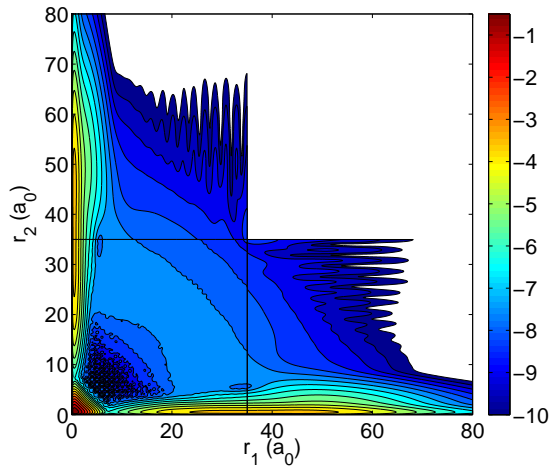}
\caption{(Color online) Two-electron wave function density as a function of the radial distance for electron 1 ($r_1$) and
2 ($r_2$) for He after time propagation for 15 cycles for a photon energy of 42.2 eV.
The laser pulse is a 14-cycle pulse with peak intensity of
10$^{14}$ W cm$^{-2}$, and a 5-cycle $\sin^2$ turn-on and turn-off. The box size is set to 35 $a_0$.
The wavefunction
density is shown on a base-10 logarithmic scale with 2 contour lines per order of magnitude.}
\label{fig:dens}
\end{figure}

In order to obtain two-photon double ionization cross sections for He, we have to analyze the wavefunction
after the initial 1s$^2$ state has been propagated over the 14-cycle laser pulse and one additional
laser field period in the absence of a laser field. Figure \ref{fig:dens} shows the
two-electron wavefunction density at the end of the calculation as a function
of the distance of both electrons from the nucleus for a photon energy of 42.2 eV.
The 1s$^2$ ground state still forms the dominant contribution to the wavefunction, 96.8\%. The total population
in the outer region is 1.9\%. Figure \ref{fig:dens} shows great
continuity between the wavefunction in the inner region and the wavefunction in the outer region.
This demonstrates that the flow between the inner and outer regions is described with very good accuracy,
both when the residual electron remains a bound electron and when the residual electron is a slower escaping
electron.

In the outer region part of Figure \ref{fig:dens}, noticeable interference structures can be seen when both
electrons are well distanced from the nucleus. These structures are quite pronounced for wavefunction densities
at the 10$^{-10}$ level, but have become substantially reduced for the contour line at a density of
3 $\times$ 10$^{-9}$. These interference structures should be expected in the
calculations. The fastest electron is allowed to go beyond the inner region radius. However, the slowest electron
is not. When this electron reaches the inner region radius, it reflects. This reflection then leads to
unphysical interference structures in the wavefunction density. These reflections are not expected to
significantly affect the double ionization cross sections, since the outer electron will continue to escape.
However, the interferences will be problematic when one is interested in more detailed properties of the
double ionization process, such as angular distributions.

The basis set used in the present calculations can, in principle, also lead to interference structures within
the inner region. The first electron in the current basis set is restricted to a maximum angular momentum
of 2. For the $(\ell_1,\ell_2) = (2,3)$ contribution to the $^1$P wavefunction, this means that the residual
ion cannot have an angular momentum exceeding 2. If the fastest electron has $\ell_2=2$, and the slow
electron $\ell_1=3$, the fast electron cannot enter the outer region, as this would result in a, presently
unallowed, residual ion with $\ell_1=3$. Figure \ref{fig:dens} shows
no significant sign of interference in the inner region. Therefore, this limitation on the wavefunction does
not appear to affect the present calculation significantly.

\begin{figure}
\includegraphics[width=8.5cm]{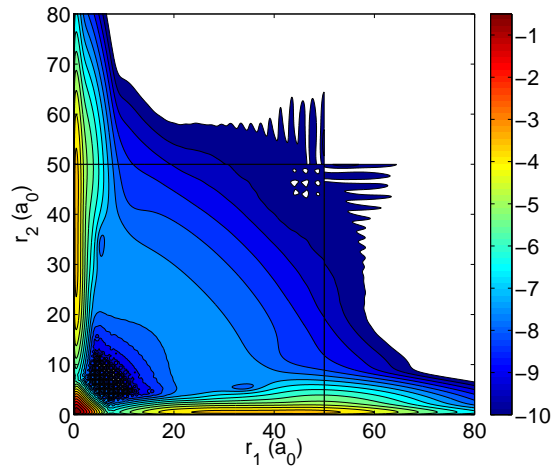}
\caption{(Color online) Two-electron wave function density as a function of the radial distance for electron 1 ($r_1$) and
2 ($r_2$) for He after time propagation for 15 cycles for a photon energy of 42.2 eV. The laser pulse is a 14-cycle pulse with peak intensity of
10$^{14}$ W cm$^{-2}$, and a 5-cycle $\sin^2$ turn-on and turn-off. The box size is set to 50 $a_0$. The wavefunction
density is shown on a base-10 logarithmic scale with 2 contour lines per order of magnitude.}
\label{fig:dens50}
\end{figure}

The effect of reflections of the slowest electron can be reduced by moving the radius of the inner-region
sphere outwards. Figure \ref{fig:dens50} shows the wavefunction density for the laser
field, but the inner region radius has been increased to 50 $a_0$. This wavefunction density is in excellent
agreement with the one obtained for an inner-region radius of 35 $a_0$. For example, the dip in the density
observed at $(r_1,r_2)$ = $(35a_0, 5a_0)$ is well reproduced in both calculations. The increase in the density of
states appears to lead to a smoother wavefunction distribution throughout the inner region, and the influence
of the boundary of the box has decreased significantly. This smoother distribution can be seen in the contour
line at $r_1, r_2 = 20 a_0$.

The calculations demonstrate a significant amount of wavefunction flow into the outer region, especially 
along the $r_1$ and $r_2$ axes. The capability to describe this single-ionization flow accurately brings
advantages to the method: we can let the two-electron emission wavepacket develop for longer when compared
to a calculation that is confined within the inner-region sphere. In a box-based calculation,
reflection of the wavefunction occurs when the first electron reaches the boundary. In the present
calculations, the first electron can enter the outer region unhindered, and it is only when the
second electron reaches the boundary that reflection of the wavefunction occurs. Therefore, we
can let the wavefunction evolve until the slowest of the two electrons reaches the boundary of the box.
Since the fastest electrons are those associated with above-threshold single ionization, the time over which
we can let the wavefunction evolve now corresponds to a time associated with double ionization, rather
than a time associated with single ionization.

A detailed comparison of the two densities in figures \ref{fig:dens} and \ref{fig:dens50} shows that the
contour lines in the inner region are smoother for a box size of 50 $a_0$ than for a box size of 35 $a_0$.
This is likely due to the finer resolution of the He$^+$ eigenfunctions for the larger box size.

In order to demonstrate the accuracy of the final-state wavefunction, we derive two-photon double
ionization cross sections from these wavefunctions. This topic has been debated extensively in the
literature (see, for example, \cite{Arg13}). In this particular calculation, our emphasis is on the
initial application of the RMT approach to double ionization processes. For the merits of the different
approaches to extract the double continuum, we refer the reader to \cite{Arg13}.
To extract the two-photon double ionization cross sections, we adopt different approaches
in the inner region and the outer region to reflect the different description
of the wavefunction in these regions. Within the inner region, we adopt an approach similar
to the one used in the investigation of triple photoionization of Li at high photon energies
\cite{Har98}. First,
we only consider the part of the wavefunction corresponding to two-electron eigenstates with a total energy
exceeding the double-ionization threshold. This part of the wavefunction is coherently transformed
back onto the product basis of He$^+$ box-based eigenstates. We then sum the population in all product states in
which both electrons are in a He$^+$ eigenstate with an energy greater than 0. To reduce the
dependence on box size, the He$^+$ eigenstate closest to the He$^+$ ionization threshold for each angular momentum
is assumed to account for both single and double ionization. The weight of its contribution to either depends
on its distance to the He$^+$ ionization threshold. Within the outer region, the double ionization yield
is obtained under the
assumption that the outer electron is in a continuum state. Channels associated with a He$^+$ state below the
He$^+$ ionization threshold then contribute to single ionization and channels associated with a He$^+$ state
above the He$^+$ ionization threshold contribute to double ionization. Once again, the He$^+$ state closest to the
ionization threshold is assumed to contribute to both single and double ionization with a weight factor
for each contribution.

Using this procedure, we have obtained two-photon double ionization cross sections for both
box sizes. For an inner-region size of 35 $a_0$, the two-photon cross section is determined to be 4.64 $\times$
10$^{-53}$ cm$^4$s, whereas for an inner-region size of 50 $a_0$, the two-photon cross section is determined to be
4.52 $\times$ 10$^{-53}$ cm$^4$s. At a box size of
50 $a_0$, only a small fraction of the wavefunction has reached the boundary of the system,  so we estimate that
the results for a box size of 35 $a_0$ overestimate the two-photon cross section by about 3-4\%. Three-photon
processes are included in the determination of the total double ionization yield. They 
contribute about 0.2-0.3\%. The three-photon contribution can be eliminated by restricting the determination
of the double ionization yield to even-parity states only. 

\begin{figure}
\includegraphics[width=8.5cm]{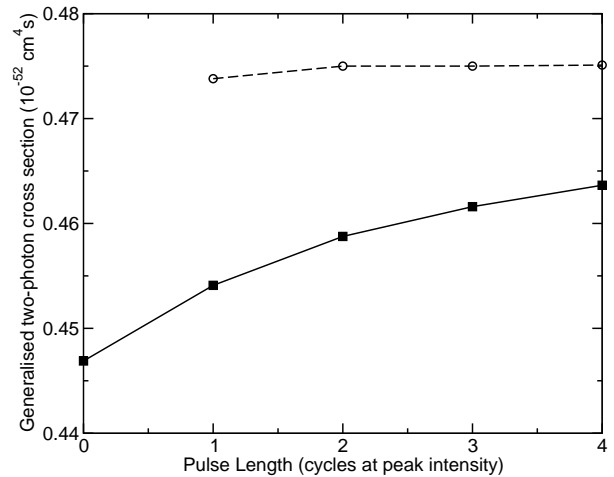}
\caption{Two-photon double ionization cross sections for He as a function of number of cycles at peak intensity
at a photon energy of 42.2 eV as obtained by the RMT approach. Cross sections derived for a pulse of
a given length are shown (full line, squares). They are compared to cross sections derived from the
difference in ionization yield obtained for a pulse of given length and for a pulse with zero
cycles at peak intensity (dashed line, circles).}
\label{fig:pulse}
\end{figure}

Figure \ref{fig:pulse} shows how the cross section, determined from the final ionization yield, depends
on the number of cycles the laser pulse is kept at peak intensity. The cross section increases from 4.47
$\times$ 10$^{-53}$ cm$^4$s for 0 cycles at peak intensity to 4.64 $\times$
10$^{-53}$ cm$^4$s for a pulse with 4 cycles at peak intensity. Alternatively, we can obtain a cross section
of 4.75 $\times$ 10$^{-53}$ cm$^4$s by examining the difference in ionization yield between a pulse with 0 cycles at
peak intensity
and a pulse with $n$ cycles at peak intensity. This latter cross section varies by 0.02\% for $n = 2,$ 3 or 4,
whereas it is about 0.25\% smaller for $n=1$.
We therefore estimate that the use of a 4-cycle pulse length underestimates the
two-photon cross section by about 3\%. Combining this with the observations for a finite box size, we
estimate the two-photon double ionization cross section to be 4.64 $\times$ 10$^{-53}$ cm$^4$s with
an accuracy of about 5-10\% within the present basis set.

\begin{figure}
\includegraphics[width=8.5cm]{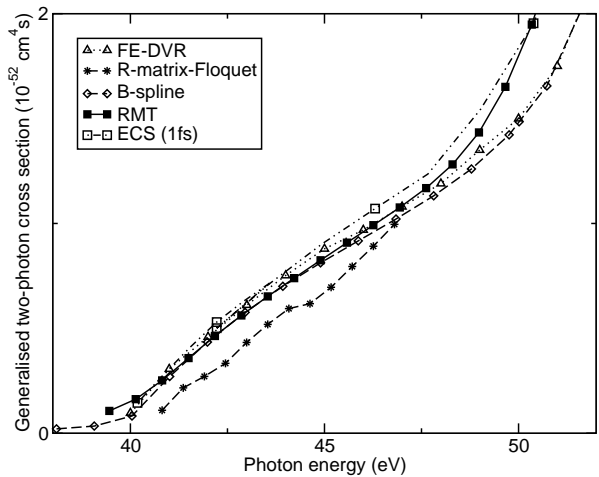}
\caption{Two-photon double ionization cross sections of He as a function of photon energy. The present cross sections
(RMT, solid squares) are compared with those obtained using the R-matrix-Floquet approach (stars) \cite{Fen03},
B-spline basis-set calculations (open diamonds) \cite{Nep10}, finite-element discrete
variable representation (FE-DVR) calculations (open triangles) \cite{Fei08}, and finite-element discrete
variable representation calculations using exterior
complex scaling (ECS) for a 1 fs pulse (open squares) \cite{Pal10}.
}
\label{fig:freq}
\end{figure}

Figure \ref{fig:freq} shows the dependence of the two-photon cross sections on photon energy, and compares the
two-photon cross sections with a selection of other available results\cite{Fen03,Fei08,Pal10,Nep10}.
The overall agreement between the cross sections
obtained through the methods presented in figure \ref{fig:freq} is very good indeed, with a spread of about
0.1 $\times$ 10$^{-52}$ cm$^4$s. The present cross sections are in particularly good agreement with the
cross sections obtained using large-scale B-spline basis sets \cite{Nep10} for photon energies between 41 and 45 eV with differences well within
1\%. This good agreement is fortuitous. As indicated above, the present cross section change by a few percent
when longer pulses and larger box sizes are used, but these changes appear to cancel each other.
The difference with the finite-element discrete-variable-representation (FE-DVR) calculations
\cite{Fei08} is typically about 5\%. 
The agreement with finite-element discrete-variable-representation employing exterior complex scaling to extract
the double continuuum (ECS) \cite{Pal10} may be better than suggested in figure \ref{fig:freq}.
The cross sections shown have been obtained for a pulse length of 1 fs, for which more data points are
available in the present range. However, cross sections obtained for a pulse length of 2 fs are smaller by about
10-20\%, which leads to a difference with the present cross sections of only 0.02 $\times$ 10$^{-52}$ cm$^4$s
at 42 and 46 eV. 

For photon energies below 41 eV, the differences are
larger: the short pulse length in the present calculation allows ionization to take place when the central
photon energy is below threshold. This effect is significantly reduced for longer pulse lengths. For
photon energies above 48 eV, the differences also become noticeable, with a difference of about 15\%
observed at a photon energy of 49 eV. In this range of photon energies, the cross sections
become affected by the rapid increase in the photoionization cross sections above 50 eV. The short pulse
length means that this increase is shifted to smaller photon energies in the present calculations. The figure
shows that in this photon energy range, the behavious of the cross sections is similar to those observed
in the ECS calculations using a similar short pulse length \cite{Pal10}. Unfortunately, it is at present
not easy to extend the calculations to longer pulse lengths. Larger box sizes are required, and as a
consequence the basis-set size increases rapidly. Further code development is needed to enable these
calculations.


The comparison with R-matrix-Floquet calculations \cite{Fen03} shows a difference in the cross sections of about 1.2 -
1.8 $\times$ 10$^{-53}$ cm$^4$s across the photon energy range considered. The origin of this difference
is likely to be the different nature of the calculations. In the Floquet calculations, only residual-ion
states below the two-photon excitation energy can be populated. In the RMT calculations, the photon energy
uncertainty still allows residual-ion states just above this energy to be excited. Therefore, some
part of the excitation spectrum may well be missed in the R-matrix-Floquet calculations, whereas that
is not the case in the RMT calculations. In addition, the R-matrix-Floquet calculations used a smaller
inner region, which reduces the resolution in the energy spectrum of the residual-ion states.

\section{Conclusions}

In conclusion, we have applied R-matrix theory including time dependence to the study of double ionization
processes in intense fields. Two-photon double ionization cross sections are obtained for He in close
agreement with other sophisticated calculations, which adopt a similar method to extract the double ionization
yield. For photon energies between 42 and 48 eV, the cross sections increase from about 0.5$\times$ 10$^{-52}$ cm$^4$s
to 1.5 $\times$ 10$^{-52}$ cm$^4$s with a spread between the different calculations of about
0.1 $\times$ 10$^{-52}$ cm$^4$s.

The present calculations demonstrate the stability of the RMT approach for the large-scale treatment of
atoms in intense fields, and demonstrate that the method can provide reliable double ionization
yields and double ionization cross sections. By combining an R-matrix inner region with an outer region, the
fundamental time constraint on the calculations is given by the time taken by the slowest electron to
reach the boundary of the inner region, instead of the fastest electron. Double ionization processes can
thus be studied on the time scale on which double ionization processes evolve rather than on the
time scale of single ionization.

The advantage of the RMT method is that it builds upon the general atomic R-matrix codes \cite{Bur11}. In
the present approach, the time-dependent approach builds upon an inner region description specific for
He. However, the approach can also be combined with a standard R-matrix approach to describe multiple
ionization for general atoms. The present results demonstrate that the RMT approach has the capability
to become a useful technique for the accurate determination of double ionization processes for general
multi-electron atoms in short light fields from first principles.

The RMT approach has been developed specifically for the treatment of general
multi-electron systems with full correlation included. At present, the time propagation of the wavefunction
is limited to a time determined by the emission of the slowest electron. To propagate the wavefunction
for longer times, a two-electron outer region needs to be combined with the present approach. The escape of
a second electron out of the inner region requires at least two electrons to be treated as continuum
electrons. In this particular application, we therefore specifically imposed that the wavefunction of the
slowest electron had to be be represented as a continuum function. The calculations demonstrate that this
change in function type poses no fundamental problem to the description of time-dependent problems. 
Adaptation of a two-electron inner-region code for He requires relatively little code development. However, 
significant changes to existing inner-region R-matrix codes are needed in order to treat general
multi-electron atoms. 

The author wishes to thank M.A. Lysaght and L.R. Moore for their substantial efforts in the development of
the RMT codes, and wishes to thank D.J. Robinson, J.S. Parker and K.T. Taylor for valuable discussions.
The author further wishes to thank M. F\o rre and A. Palacios for providing cross sections in numerical form.
This research
was sponsored by the Engineering and Physical Sciences Research Council (UK) under grant ref. no. G/055416/1.
This work made use of the facilities of HECToR, the UK's national high-performance computing service,
which is provided by UoE HPCx Ltd at the University of Edinburgh, Cray Inc and NAG Ltd, and funded by
the Office of Science and Technology through EPSRC's High End Computing Programme.


\begin{thebibliography}{}
\bibitem{Yab13} M. Yabashi \textit{et al}, \textit{J. Phys. B} {\bf 46}, 164001 (2013)
\bibitem{Feld13} J. Feldhaus \textit{et al}, \textit{J. Phys. B} {\bf 46}, 164002 (2013)
\bibitem{Bos13} C. Bostedt \textit{et al}, \textit{J. Phys.B} {\bf 46}, 164003
(2013).
\bibitem{Ric09} M. Richter \textit{et al}, \textit{Phys. Rev. Lett.} {\bf 102}, 163002 (2009)
\bibitem{Rud12} B. Rudek \textit{et al}, \textit{Nature Photonics} {\bf 6}, 858 (2012)
\bibitem{Kar13} E.T. Karamatskos, D. Markellos and P. Lambropoulos, \textit{J. Phys. B} {\bf 46}, 164011 (2013)
\bibitem{Col02} J. Colgan and M.S. Pindzola, \textit{Phys. Rev. Lett.} {\bf 88}, 173002 (2002)
\bibitem{Fen03} L. Feng and H.W. van der Hart, \textit{J. Phys. B} {\bf 36}, L1 (2003)
\bibitem{Fei08} J. Feist, S. Nagele, R. Pazourek, E. Persson, B.I. Schneider, L.A. Collins
and J. Burgd\"orfer, \textit{Phys. Rev. A} {\bf 77}, 043420 (2008)
\bibitem{Gua09} X.X. Guan, O. Zatsarinny, C.J. Noble, K. Bartschat and B.I. Schneider, \textit{J. Phys. B}
{\bf 42}, 134015 (2009)
\bibitem{Pal10} A. Palacios, D.A. Horner, T.N. Rescigno and C.W. McCurdy, \textit{J. Phys. B} {\bf 43},
194003 (2010)
\bibitem{Nep10} R. Nepstad, T. Birkeland and M. F\o rre, \textit{Phys. Rev. A} {\bf 81}, 063402 (2010)
\bibitem{Arg13} L. Argenti, R. Pazourek, J. Feist, S. Nagele, M. Liertzer, E. Persson, J. Burgd\"orfer,
and E. Lindroth, \textit{Phys. Rev. A} {\bf 87}, 053405 (2013)
\bibitem{Yip13} F.L. Yip, T.N. Rescigno, C.W. McCurdy and F. Mart\'\i n, \textit{Phys. Rev. Lett.} {\bf 110},
173001 (2013)
\bibitem{Har07} H.W. van der Hart, M.A. Lysaght, and P.G. Burke,
\textit{Phys. Rev. A} {\bf 76}, 043405 (2007)
\bibitem{Nik08} L.A.A. Nikolopoulos, J.S. Parker and K.T. Taylor, \textit{Phys. Rev. A} {\bf 78},
063420 (2008) 
\bibitem{Lys09} M. A. Lysaght, H. W. van der Hart, and P. G. Burke, \textit{Phys. Rev. A}
{\bf 79}, 053411 (2009)
\bibitem{Lys11} M.A. Lysaght, L.R. Moore, L.A.A. Nikolopoulos, J.S. Parker,
H.W. van der Hart and K.T. Taylor, \textit{Quantum Dynamic Imaging:
  Theoretical and Numerical Methods} (eds. A.D. Bandrauk and M. Yu. Ivanov, Springer:NewYork, 2011)
  107-134
\bibitem{Moo11} L.R. Moore, M.A. Lysaght, L.A.A. Nikolopoulos, J.S. Parker,
H.W. van der Hart and K.T. Taylor, \textit{J. Mod. Opt.} {\bf 58}, 1132 (2011)
\bibitem{Lys09b} M.A. Lysaght, P.G. Burke and H.W. van der Hart, \textit{Phys. Rev. Lett.} {\bf 102},
193001 (2009)
\bibitem{Moo11b} L.R. Moore, M.A. Lysaght, J.S. Parker, H.W. van der Hart
and K.T. Taylor,  \textit{Phys. Rev. A} {\bf 84}, 061404 (2011)
\bibitem{Bro12} A.C. Brown, S. Hutchinson, M.A. Lysaght, and H.W. van der Hart,
\textit{Phys. Rev. Lett.} {\bf 108}, 063006 (2012)
\bibitem{Har97} H.W. van der Hart, \textit{J. Phys. B} {\bf 30}, 453 (1997)
\bibitem{Sco12} M.P. Scott, A.J. Kinnen and M.W. McIntyre, \textit{Phys. Rev. A} {\bf 86}, 032707 (2012)
\bibitem{Smy98} E.S. Smyth, J.S. Parker and K.T. Taylor, \textit{Comp. Phys. Comm.} {\bf 114}, 1 (1998)
\bibitem{Har98} H.W. van der Hart and C.H. Greene, \textit{Phys. Rev. Lett.} {\bf 81}, 4333 (1998)
\bibitem{Bur11} P. G. Burke, \emph{R-Matrix Theory of Atomic Collisions},
(Springer Verlag, Heidelberg, 2011)

\end{thebibliography}
\end{document}